\documentclass[preprint,showpacs,pra,a4paper]{revtex4}
\usepackage{graphicx}

\usepackage{graphicx}

\begin{document}

\preprint{Special issue on quantum electrodynamics}

\title{Theory of non-Markovian decay of a cascade atom in high-Q cavities
and photonic band-gap materials}
\author{B.M.~Garraway}
 \affiliation{Department of Physics and Astronomy,\\
 University of Sussex, Falmer, Brighton, BN1 9QH, United Kingdom}

\author{B.J.~Dalton}
 \affiliation{ARC Centre of Excellence for Quantum-Atom Optics,\\
 and Centre for Atom Optics and Ultrafast Spectroscopy,\\
 Swinburne University of Technology, Melbourne, Victoria 3122, Australia}

\date{\today}

\begin{abstract}

The dynamics of a three-level atom in a cascade configuration with both
transitions coupled to a single structured reservoir of quantized field
modes is treated using Laplace transform methods applied to the coupled
amplitude equations. Results are also obtained from master equations by two
different approaches, that is, involving either pseudomodes or quasimodes.
Two different types of reservoir are considered, namely a high-Q cavity and
a photonic band-gap system, in which the respective reservoir structure
functions involve Lorentzians. Non-resonant transitions are included in the
model. In all cases non-Markovian behaviour for the atomic system can be
found, such as oscillatory decay for the high-Q cavity case and population
trapping for the photonic band-gap case. In the master equation approaches,
the atomic system is augmented by a small number of pseudomodes or
quasimodes, which in the quasimode approach themselves undergo Markovian
relaxation into a flat reservoir of continuum quasimodes.  Results from
these methods are found to be identical to those from the Laplace transform
method including two-photon excitation of the reservoir with both emitting
sequences. This shows that complicated non-Markovian decays of an atomic
system into structured EM field reservoirs can be described by Markovian
models for the atomic system coupled to a small number of pseudomodes or
quasimodes.
\end{abstract}


\maketitle

\section{INTRODUCTION}

\label{sec:Intro}

The study of open quantum systems, in which the interaction of the quantum
system with the environment is taken into account, is of fundamental
importance in several areas of physics. One such area is quantum optics \cite%
{Scully97a}, where the quantum system may be an atom (or an atom combined
with a single mode laser field) and the environment with which the atom
interacts may be a continuum of quantised modes of the electromagnetic field
(or a set of vibrational modes in a solid). Another area is quantum
measurement theory \cite{Auletta01a,Zurek03a}, where the environment
includes the macroscopic measuring apparatus, whose states ``record'' the
results of measurements on the quantum system itself. Explaining the
emergence of the classical world \cite{Zurek93a,Zurek03a} is related to the
way the presence of the environment destroys coherence between certain
states of macroscopic systems, resulting in only the ``pointer states''
remaining stable---any quantum superpositions of pointer states are rapidly
converted into mixed states. A third area is that of degenerate quantum
gases \cite{Pitaevskii03a}, where the quantum system is an atomic
Bose-Einstein condensate or atomic Fermi gas, and the environment may
consist of those atoms in thermally excited states, as distinct from being
in the macroscopically occupied condensate state or in states within the
Fermi surface. As in some of these examples, the quantum system itself may
be microscopic or macroscopic.

In many cases the interaction between the quantum system and the environment
(also referred to as the bath or reservoir) involves coupling constants and
reservoir mode densities that have a slowly varying frequency dependence. In
such cases the reservoir structure function (which is the product of the
mode density with the square of the coupling constant) is also slowly
varying, resulting in the reservoir correlation time (which is the inverse
of the bandwidth for the reservoir structure function) being very short
compared to the time scales for dynamic evolution of the quantum system.
This situation enables the dynamical behaviour of the quantum system and its
interaction with the environment to be described by Markovian master
equations (see e.g.\ \cite{Scully97a}). A large literature exists where the
dynamical behaviour of quantum systems coupled to the environment has been
successfully explained via Markovian master equations and related methods,
and indeed much of quantum optics (see \cite{Scully97a,Barnett97a}) falls
into this category.

In recent years however, there has been an interest in open quantum systems
where the conditions required for Markovian behaviour do not necessarily
apply. Cases include atom lasers \cite{Hope00a}, quantum Brownian motion 
\cite{Strunz04a,Breuer01a}, systems with conditioned evolution (such as
associated with photodetection) \cite{Wiseman03a}, decoherence in large
scale quantum computers \cite{Privman04a,DiVincenzo03a,Dalton03b} and in
macroscopic systems generally \cite{Braun01a,Strunz03a,Strunz03b}. A further
non-Markovian situation occurs for atomic systems coupled to structured
reservoirs of electromagnetic (EM) field modes, where either the coupling
constants or the mode density (or both) change rapidly with frequency. This
situation can occur for atoms in high-Q cavities or in photonic band-gap
(PBG) systems, and a general review of such situations is given by
Lambropoulos et al.\ \cite{Lambropoulos00a}.

A number of methods for treating non-Markovian problems have been
formulated. These include: non-Markovian master equations \cite%
{Zwanzig64a,Nakajima58a,Dalton82a,Barnett01a}; the time-convolutionless
projection operator master equation \cite{Shibata77a,Strunz04a,Breuer01a};
Heisenberg equations of motion \cite{Cresser00a}; stochastic wave function
methods for non-Markovian processes \cite%
{Breuer99a,Strunz99a,Walls99a,Molmer99a,John97a,Imamoglu96a,Strunz04a,Wiseman03a,Breuer04a}%
; methods based on the essential states approximation or resolvent operators 
\cite{Lambropoulos00a,Knight99a,Lambropoulos97a,Law00a}; the pseudomode
approach \cite{Garraway97a,Garraway97b}; Fano diagonalisation \cite%
{Fano61a,Barnett00a}; and various short time scale methods \cite%
{Braun01a,Strunz03a,Privman02a,Privman03a,Privman04b,Duan97a,Zagury01a}. The
last four approaches are less powerful in the formal sense, but often
simpler to apply and interpret.

In recent work (\cite{Dalton01a,Dalton01b,Dalton02a,Dalton03a}) we have
studied non-Markovian processes for atomic systems coupled to a structured
reservoirs of EM field modes, with applications to high-Q cavities and PBG
systems \cite{Lambropoulos00a}. Our approach has been based on the essential
states approximation, Fano diagonalisation and pseudomodes. In Ref.\cite%
{Dalton03a} the case of a three level \emph{cascade }(or ladder) system
coupled to a structured reservoir of EM modes was treated using the
essential states approach, following a method similar to \cite{Law00a}. The
general equations for the coupled amplitudes and their Laplace transforms
were obtained, and the resulting integral equation solved via numerical
methods, based on discretising frequency space to give a matrix equation
that is equivalent to the original integral equation and also utilising
analytical continuation in the complex $s$ plane. The decay of the initially
excited upper state was found as a function of time. Our application in \cite%
{Dalton03a} was restricted to the case where the structured reservoir was
due to a high-Q cavity. In the present paper, we apply the same method also
to the case where the structured reservoir is due to a model photonic
band-gap. In addition some further high-Q cavity cases are examined, where
the cavity resonance is detuned from the atomic transition frequency. Other
non-Markovian studies of the cascade system in a PBG system have been
carried out in Refs.\ \cite{Nikolopoulos00a,Bay98a}. In these papers a
discontinuous mode density is used, whereas we represent the PBG by a
difference of two Lorentzian functions. Quantitative comparisons between
their results and ours are therefore not possible, only general features
such as population trapping effects are in common. For completeness and to
set out the notation, we include here in Section \ref{sec:Essential States
Theory} the key integral equations for the cascade system decay, based on
the essential states approach, that were derived in \cite{Dalton03a}.
Results for both the high-Q cavity and PBG cases are presented in Section %
\ref{sec: Applications}. The dynamics is also treated via master equations
by two different approaches, that is, involving either pseudomodes or
quasimodes. In the master equation approach, the atomic system is augmented
by a small number of pseudomodes or quasimodes, which in the quasimode
approach themselves undergo Markovian relaxation into a flat reservoir of
continuum quasimodes. The master equation theory is presented in Section \ref%
{sec:ME}, with the pseudomode approach \cite{Garraway97a,Garraway97b} being
described in Subsection \ref{sec:Pseudomode}, and a treatment based on
quasimodes \cite{Dalton01a,Dalton01b} given in Subsection \ref{sec:Quasimode}%
. Results from the master equation method are compared to those from the
Laplace transform method. A summary of the paper is set out in Section \ref%
{sec:Summary}. 

\section{ESSENTIAL STATES THEORY}

\label{sec:Essential States Theory}

\subsection{Model System}

\begin{figure}
 \includegraphics[width=3cm]{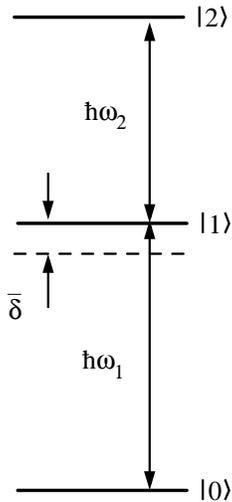}
\caption[fig1]{Energy levels and energy differences in the three level
system. The off-set $\bar{\protect\delta}$ represents the difference between
the energy of level 1 and the midpoint between levels 0 and 2. }
\label{Fig1}
\end{figure}

The system consists of a three level cascade atom, shown in figure \ref{Fig1}%
, coupled to a bath (or reservoir) of bosonic modes. The bath modes would
usually be associated with the quantum EM field, but other bosonic baths
(such as lattice vibrations in a solid) might realise a similar model. For
convenience, we will refer to the bath quanta as photons, but other quanta
could be involved (such as phonons in the lattice vibration case).

The atomic states are illustrated in figure \ref{Fig1}: the energy
difference between states 1 and 0 is $\hbar \omega _1$, and between 2 and 1
the difference is $\hbar \omega _2$. For later use we also define a
frequency offset $\bar{\delta}$, such that 
\begin{equation}
\bar{\delta} = ( \omega_1 - \omega_2 ) /2 .  \label{def.deltabar}
\end{equation}
The offset determines the energy difference between the two transitions. If
the offset is large, we would expect the system to behave as though there
are two independent reservoirs. However, for relatively small offsets a
photon may well be emitted from one transition, and later re-absorbed on the
other transition. The bath modes have frequencies $\omega _{\lambda }$,
raising and lowering operators $\hat{a}_{\lambda }^{\dagger }$ and $\hat{a}%
_{\lambda }$ and the density of modes is $\rho _{\lambda }$. Frequency
dependent coupling constants $g_{\lambda 1}$ and $g_{\lambda 2}$ specify the
coupling between atomic transitions $0-1$ and $1-2$ and the
reservoir.

The Hamiltonian for the system in the rotating wave approximation is given
by: 
\begin{eqnarray}
\hat{H} &=&\hbar \Biggl[\omega _{1}|1\rangle \langle 1|+(\omega _{1}+\omega
_{2})|2\rangle \langle 2|+\sum_{\lambda }\omega _{\lambda }\hat{a}_{\lambda
}^{\dagger }\hat{a}_{\lambda }  \nonumber \\
&&+\sum_{\lambda }\left[ g_{\lambda 1}\left( \hat{a}_{\lambda }^{\dagger
}|0\rangle \langle 1|+\hat{a}_{\lambda }|1\rangle \langle 0|\right)
+g_{\lambda 2}\left( \hat{a}_{\lambda }^{\dagger }|1\rangle \langle 2|+\hat{a%
}_{\lambda }|2\rangle \langle 1|\right) \right] \Biggr] .  \label{eq:hamil}
\end{eqnarray}%
Initially the atom is assumed to be in the upper state $|2\rangle $ and the
bath modes are empty of photons, so the initial state vector is 
\begin{equation}
\left\vert \Psi (0)\right\rangle =|2\rangle |0_{\lambda }\rangle .
\label{eq. initial state}
\end{equation}%
The atom plus bath system evolves according to the Schr\"{o}dinger equation.
The initial state assumed together with the rotating wave approximation
allows for the atomic state $|1\rangle $ to be coupled to one photon bath
states and atomic state $|0\rangle $ to be coupled to bath states with two
photons---the latter may be associated with the same mode or two different
modes.

\subsection{Coupled Amplitude Equations}

The time evolving state vector is written in the form: 
\begin{eqnarray}
\left\vert \Psi (t)\right\rangle &=&c_{2}e^{-i(\omega _{1}+\omega
_{2})t}|2\rangle |0_{\lambda }\rangle +\sum_{\lambda }c_{1\lambda
}e^{-i(\omega _{1}+\omega _{\lambda })t}|1\rangle |1_{\lambda }\rangle 
\nonumber \\*  
&&\!\!\!+\sum_{\lambda }c_{0\lambda \lambda }e^{-2i\omega _{\lambda
}t}|0\rangle |2_{\lambda }\rangle +\sum_{\lambda ,\mu \,,\lambda <\mu
}c_{0\lambda \mu }e^{-i(\omega _{\lambda }+\omega _{\mu })t}|0\rangle
|1_{\lambda }1_{\mu }\rangle ,  \label{eq. state vector}
\end{eqnarray}%
where $c_{2},c_{1\lambda },c_{0\lambda \lambda }$ and $c_{0\lambda \mu }$
are interaction picture amplitudes for the various states. A
straight-forward substitution into the Schr\"{o}dinger equation leads to a
set of coupled equations for these amplitudes.

The overall aim is to find the evolution of the atomic initial state, e.g.\ $%
c_{2}(t)$. We first obtain a set of coupled equations for the Laplace
transforms of the amplitude equations, and then apply the approach of \cite%
{Law00a} to make a convenient change of the amplitudes by incorporating the
coupling constants via $\bar{c}_{2}=\bar{b}_{2},\bar{c}_{1\lambda
}=g_{\lambda 2}\bar{b}_{1\lambda },\bar{c}_{0\lambda \mu }=g_{\lambda
2}g_{\mu 1}\bar{b}_{0\lambda \mu }\quad (\lambda <\mu )$ and $\bar{c}%
_{0\lambda \lambda }=g_{\lambda 2}g_{\lambda 1}\bar{b}_{0\lambda \lambda }$.
Laplace transforms are denoted by a bar, and associated with the complex
variable $s$. The coupled equations obtained are \cite{Dalton03a}:%
\begin{eqnarray}
s\bar{b}_{2}(s)-1 &=&-i\sum_{\lambda }g_{\lambda 2}^{2}\bar{b}_{1\lambda
}(s+i(\omega _{\lambda }-\omega _{2}))  \label{eq. LT1} \\
s\bar{b}_{1\lambda }(s) &=&-i\sum_{\mu ,(\mu >\lambda )}g_{\mu 1}^{2}\bar{b}%
_{0\lambda \mu }(s+i(\omega _{\mu }-\omega _{1}))  \nonumber \\
&&-i\sum_{\mu ,(\mu <\lambda )}g_{\mu 1}^{2}\alpha _{\lambda \mu }\bar{b}%
_{0\mu \lambda }(s+i(\omega _{\mu }-\omega _{1}))  \nonumber \\
&&-ig_{\lambda 1}^{2}\sqrt{2}\bar{b}_{0\lambda \lambda }(s+i(\omega
_{\lambda }-\omega _{1}))-i\bar{b}_{2}(s+i(\omega _{2}-\omega _{\lambda }))
\label{eq. LT2} \\
s\bar{b}_{0\lambda \lambda }(s) &=&-i\sqrt{2}\bar{b}_{1\lambda }(s+i(\omega
_{1}-\omega _{\lambda }))  \label{eq. LT3} \\
s\bar{b}_{0\lambda \mu }(s) &=&-i\bar{b}_{1\lambda }(s+i(\omega _{1}-\omega
_{\mu }))-i\alpha _{\lambda \mu }\bar{b}_{1\mu }(s+i(\omega _{1}-\omega
_{\lambda })).  \label{eq. LT4}
\end{eqnarray}%
These involve a frequency independent parameter: 
\begin{equation}
\alpha _{\lambda \mu }=\frac{g_{\lambda 1}g_{\mu 2}}{g_{\lambda 2}g_{\mu 1}}
.  \label{eq. alpha}
\end{equation}%

It is not possible to obtain explicit solutions for the new Laplace
transform amplitudes. However, we can proceed further by first eliminating $%
\bar{b}_{0\lambda \mu },\bar{b}_{0\lambda \lambda }$ in the above equations,
leading to a set of coupled equations for the remaining amplitudes $\bar{b}%
_{2}(s)$ and $\bar{b}_{1\mu }(s)$.

\subsection{Integral Equation}

We then eliminate $\bar{b}_{2}(s)$ in favour of $\bar{b}_{1\mu }(s)$ (though
not the reverse) giving a set of equations (see \cite{Dalton03a}) for the
remaining amplitude $\bar{b}_{1\mu }(s)$:%
\begin{eqnarray}
-i/s &=&\left( s+i(\omega _{\lambda }-\omega _{2})+\sum_{\eta }\frac{g_{\eta
1}^{2}}{s+i(\omega _{\lambda }+\omega _{\eta }-\omega _{1}-\omega _{2})}%
\right) \bar{b}_{1\lambda }(s+i(\omega _{\lambda }-\omega _{2})  \nonumber \\
&&+\sum_{\mu }\left( g_{\mu 1}^{2}\frac{g_{\lambda 1}g_{\mu 2}}{g_{\lambda
2}g_{\mu 1}}\frac{1}{s+i(\omega _{\lambda }+\omega _{\mu }-\omega
_{1}-\omega _{2})}+\frac{g_{\mu 2}^{2}}{s}\right) \bar{b}_{1\mu }(s+i(\omega
_{\mu }-\omega _{2})) .  \nonumber \\
&&  \label{eq. LT-b1lambda}
\end{eqnarray}%
The above equation is for the amplitude $\bar{b}_{1\lambda }(s)$ ---it is of
course coupled to all similar amplitudes $\bar{b}_{1\mu }(s)$.

By converting the sums to integrals, i.e.\ $\sum_{\mu }\longrightarrow \int
d\omega _{\mu }\rho (\omega _{\mu })$, we obtain a Fredholm integral
equation of the second kind \cite{Dalton03a}%
\begin{equation}
\overline{f}(\omega _{\lambda })+\int d\omega _{\mu }\,K(\omega _{\lambda
},\omega _{\mu })\,\overline{f}(\omega _{\mu })=d(\omega _{\lambda })
\label{eq. Basic Int Eqn}
\end{equation}%
with 
\begin{eqnarray}
\overline{f}(\omega _{\lambda }) &=&\bar{b}_{1\lambda }(s+i(\omega _{\lambda
}-\omega _{2}))  \label{eq. f} \\
K(\omega _{\lambda },\omega _{\mu }) &=&B(\omega _{\lambda },\omega _{\mu
})/A(\omega _{\lambda })  \label{eq. K} \\
d(\omega _{\lambda }) &=&C/A(\omega _{\lambda })  \label{eq. D} \\
A(\omega _{\lambda }) &=&s+i(\omega _{\lambda }-\omega _{2})+\int d\omega
_{\eta }\rho (\omega _{\eta })\frac{g_{\eta 1}^{2}}{s+i(\omega _{\lambda
}+\omega _{\eta }-\omega _{1}-\omega _{2})}  \label{eq. A} \\
B(\omega _{\lambda },\omega _{\mu }) &=&\rho (\omega _{\mu })\left( g_{\mu
1}^{2}\frac{g_{\lambda 1}g_{\mu 2}}{g_{\lambda 2}g_{\mu 1}}\frac{1}{%
s+i(\omega _{\lambda }+\omega _{\mu }-\omega _{1}-\omega _{2})}+\frac{g_{\mu
2}^{2}}{s}\right)  \label{eq. B} \\
C &=&-i/s  \label{eq. C}
\end{eqnarray}%
The quantities $\overline{f}(\omega _{\lambda }),K(\omega _{\lambda },\omega
_{\mu }),A(\omega _{\lambda }),B(\omega _{\lambda },\omega _{\mu }),C$ and $%
d(\omega _{\lambda })$ are all functions of the Laplace variable $s$. The
quantity $K$ performs the role of the kernel for the integral equation for
the quantity $\overline{f}$, and the quantity $d$ makes the integral
equation inhomogeneous.

As pointed out in \cite{Dalton03a}, the integral equations for $\bar{b}%
_{1\lambda }(s)$ and the other related equations for $\bar{b}_{2}(s)$ and $%
\bar{b}_{0\lambda \mu }(s)$ only involve the reservoir quantities $%
g_{\lambda 1},g_{\lambda 2},\rho _{\lambda }$ in terms of \emph{reservoir
structure functions} 
\begin{eqnarray}
R_{1}(\omega _{\lambda }) &=&\rho (\omega _{\lambda })g_{\lambda 1}^{2}
\label{eq. Res Fn1} \\
R_{2}(\omega _{\lambda }) &=&\rho (\omega _{\lambda })g_{\lambda 2}^{2}
\label{eq. Res Fn2}
\end{eqnarray}

Since the atomic density operator%
\begin{equation}
\widehat{\rho }_{A}=Tr_{F}\,\left\vert \Psi \right\rangle \langle \Psi |
\label{eq. Atom Dens Opr}
\end{equation}%
also only involves reservoir structure functions and the reduced amplitudes $%
b_{2}(t),b_{1\lambda }(t)$ and $b_{0\lambda \mu }(t)$, an important result
follows that the overall dynamics is entirely determined by the reservoir
structure functions.

The integral equation (\ref{eq. Basic Int Eqn}) can be solved in different
ways, such as: (a) numerical methods in which the integral equation is
converted to a matrix equation (see next sections); (b) expansions using
bi-orthogonal eigenfunctions (see Ref. \cite{Dalton03a}); (c) expansions
such as the Fredholm expansion (see textbooks on integral equations).

\subsection{Numerical Solution of Integral Equation}

\label{sec:num_integral}

By discretising the mode frequencies $\omega _{\lambda }$ the integral
equation (\ref{eq. Basic Int Eqn}) can be converted to a matrix equation
(the rows and columns are specified by the discrete frequencies and the
integral over $\omega _{\mu }$ is approximated by a discrete sum) of the
form 
\begin{equation}
(\mathbf{K}+\mathbf{I})\overline{\mathbf{f}}=\mathbf{d,}
\label{eq. Matrix Basic Int Eqn}
\end{equation}%
which we then can solve for $\overline{\mathbf{f}}$.

However, the obvious simple solution of the form $\overline{\mathbf{f}}=(%
\mathbf{K}+\mathbf{I})^{-1}\mathbf{d}$ is not correct. To get the time
evolution of $b_{2}$ for example, we need both its real and imaginary parts, 
$b_{2r}$ and $b_{2i}$. This means we need the separate Laplace transforms $%
\bar{b}_{2r}(s)$ and $\bar{b}_{2i}(s)$ to invert, and this in turn requires
us to separately obtain $\overline{f_{r}}(s)$ and $\overline{f_{i}}(s)$.
Thus we need the \emph{separate} real and imaginary parts of $\overline{f}$,
both being functions of the complex Laplace variable $s$. These separate
parts are obtained by \emph{analytic continuation} of the real and imaginary
parts of $K(\omega _{\lambda },\omega _{\mu })$, $\overline{f}(\omega
_{\lambda })$ and $D(\omega _{\lambda })$ on the real $s$-axis. In other
words, we first break the equation $(\mathbf{K}+\mathbf{I})\overline{\mathbf{%
f}}=\mathbf{d}$ into its real and imaginary parts on the basis of the
variable $s$ being \emph{real}, solve for $\overline{f_{r}}(s)$ and $%
\overline{f_{i}}(s)$ with real $s$, then obtain $\overline{f_{r}}(s)$ and $%
\overline{f_{i}}(s)$ for complex $s$ by analytic continuation.

In matrix form in terms of the real and imaginary parts, the integral
equation (\ref{eq. Matrix Basic Int Eqn}) becomes 
\begin{equation}
\left( 
\begin{array}{c|c}
\mathbf{K}_{r}+\mathbf{I} & -\mathbf{K}_{i} \\ \hline
\mathbf{K}_{i} & \mathbf{K}_{r}+\mathbf{I}%
\end{array}%
\right) \left( 
\begin{array}{c}
\overline{\mathbf{f}_{r}} \\ \hline
\overline{\mathbf{f}_{i}}%
\end{array}%
\right) =\left( 
\begin{array}{c}
\mathbf{d}_{r} \\ \hline
\mathbf{d}_{i}%
\end{array}%
\right) .  \label{eq. Real Imag Compts}
\end{equation}%
As emphasised above, the real, imaginary parts are denoted $\mathbf{K}_{r}$, 
$\mathbf{K}_{i}$ etc. and are identified for $s$ real. Analytic continuation
enables solution for $\overline{\mathbf{f}_{r}}$ and $\overline{\mathbf{f}%
_{i}}$ which will apply for \emph{all} $s$.

Solving (\ref{eq. Real Imag Compts}) for $\overline{\mathbf{f}_{r}}$ and $%
\overline{\mathbf{f}_{i}}$ by matrix inversion, gives the LT amplitude $\bar{%
b}_{1\lambda }(s+i(\omega _{\lambda }-\omega _{2}))$. The coupled amplitude
equations (\ref{eq. LT1}) give $\bar{b}_{2}(s)$. For the discretised
frequency form, the real and imaginary parts can be written as the scalar
products \cite{Dalton03a} 
\begin{eqnarray}
\bar{b}_{2r}(s) &=&(1+\mathbf{r}\cdot \overline{\mathbf{f}_{i}})/s
\label{eq. LT b real} \\
\bar{b}_{2i}(s) &=&-\mathbf{r}\cdot \overline{\mathbf{f}_{r}}/s,
\label{eq. LT b imag}
\end{eqnarray}%
where $\mathbf{r}\equiv \{\rho _{\lambda }g_{\lambda 2}^{2}\}$. The
amplitude $b_{2}(t)$ then follows from a numerical inverse Laplace transform
of \emph{both} $\bar{b}_{2r}(s)$ and $\bar{b}_{2i}(s)$.

\section{APPLICATIONS}

\label{sec: Applications}

\subsection{Lorentzian resonances}

Before looking at two special cases, we will first generate some results for
a reservoir structure composed of a sum of Lorentzian resonances. The
resonances need not be well separated from each other in frequency space,
but must be well above zero frequency. The general form of structure
function is then: 
\begin{equation}
\rho g_{1}^{2}(\omega )=\eta \rho g_{2}^{2}(\omega )=\sum_{\alpha }Z_{\alpha
}\,\cdot \,\frac{\Gamma _{\alpha }}{2\pi }\,\cdot \,\frac{1}{(\omega -\omega
_{\alpha }^{c})^{2}+(\Gamma _{\alpha }/2)^{2}},  \label{lorentz.sum}
\end{equation}%
where a simple scaling $\eta $ has been introduced between the strength of
coupling for the upper and lower transitions, $g_{\lambda 1}^{2}=\eta
g_{\lambda 2}^{2}$. In this situation $\alpha _{\lambda \mu }=1$, [see
equation (\ref{eq. alpha})]. The $\alpha $th Lorentzian is weighted by $%
Z_{\alpha }$ (not necessarily positive) and has a resonance centred at the
frequency $\omega _{\alpha }^{c}$ with width $\Gamma _{\alpha }$.

For these cases of reservoir structure function, the integral in equation (%
\ref{eq. A}) is straightforward and gives for the function $A$ 
\begin{equation}
A(\omega _{\lambda })=s+i(\omega _{\lambda }-\omega _{2})+\sum_{\alpha }%
\frac{Z_{\alpha }}{s+\Gamma _{\alpha }/2+i(\omega _{\lambda }+\omega
_{\alpha }^{c}-\omega _{1}-\omega _{2})} 
 . 
\end{equation}

In what follows we define the reservoir frequency and resonances relative to
the average of the transition frequencies, i.e.\ we let 
\begin{eqnarray}
\Delta \omega &=&\omega _{\lambda }-(\omega _{1}+\omega _{2})/2  \nonumber \\
\Delta \omega ^{\prime } &=&\omega _{\mu }-(\omega _{1}+\omega _{2})/2
\end{eqnarray}%
and define 
\begin{equation}
\delta _{\alpha }=\omega _{\alpha }^{c}-(\omega _{1}+\omega _{2})/2.
\end{equation}%
The detuning $\delta _{\alpha }$ is then the offset of the $\alpha $th
Lorentzian from the average frequency such that if $\delta _{\alpha }=\pm 
\bar{\delta}$ the $\alpha $th Lorentzian is resonant with $\omega _{1}$, or $%
\omega _{2}$, according to the sign of $\bar{\delta}$. Then, together with
equation (\ref{def.deltabar}) for $\bar{\delta}$, we find in this notation
that 
\begin{equation}
A(\Delta \omega )=s+i(\Delta \omega +\bar{\delta})+\sum_{\alpha }\frac{%
Z_{\alpha }}{s+\Gamma _{\alpha }/2+i(\Delta \omega +\delta _{\alpha })}
\end{equation}%
and 
\begin{equation}
B(\Delta \omega ,\Delta \omega ^{\prime })=\left[ \frac{1}{\eta s}+\frac{1}{%
s+i(\Delta \omega +\Delta \omega ^{\prime })}\right] \sum_{\alpha }Z_{\alpha
}\,\cdot \,\frac{\Gamma _{\alpha }}{2\pi }\,\cdot \,\frac{1}{(\Delta \omega
^{\prime }-\delta _{\alpha })^{2}+(\Gamma _{\alpha }/2)^{2}}.
\end{equation}%
As before, the kernel is then $K(\Delta \omega ,\Delta \omega ^{\prime
})=B(\Delta \omega ,\Delta \omega ^{\prime })/A(\Delta \omega )$.

\subsection{Case A: High-Q Cavity Model}

\label{sec:hiq}

In this section we consider the special case of the three-level atom coupled
to a \emph{single high-Q cavity mode}. For simplicity, identical coupling
constants will be assumed. The atomic transition frequencies may be unequal.
Thus we have $g_{\lambda 1}=g_{\lambda 2}=g_{\lambda }$, and $\eta =1$ in
equation (\ref{lorentz.sum}). In this case the frequency dependence of the
reservoir structure function will be due to resonant behaviour of the \emph{%
coupling constants}, as a quasimode theory of such systems demonstrates \cite%
{Dalton02a}. Thus, parameterising the single weight $Z_{1}$ by $\Omega ^{2}$
we have: 
\begin{eqnarray}
R_{1} &=&R_{2}=R=\rho _{\lambda }g_{\lambda }^{2}  \label{eq. RSF High Q 1}
\\
&=&\frac{\Gamma \Omega ^{2}}{2\pi }\,\cdot \,\frac{1}{(\Delta \omega -\delta
)^{2}+(\Gamma /2)^{2}},  \label{eq. RSF High Q 2}
\end{eqnarray}%
where we denote the single detuning $\delta _{1}$ by $\delta $.

The kernel $K$ [see equation (\ref{eq. K})] is given by:\newline
\begin{equation}
K(\Delta \omega ,\Delta \omega ^{\prime })=\frac{\Gamma \Omega ^{2}}{2\pi }%
\frac{(s+i(\Delta \omega +\delta )+\Gamma /2)(2s+i(\Delta \omega +\Delta
\omega ^{\prime })}{s((\Delta \omega ^{\prime }-\delta )^{2}+(\Gamma
/2)^{2})(s+i(\Delta \omega +\Delta \omega ^{\prime }))Q(\Delta \omega )},
\label{eq. K High Q}
\end{equation}%
where $Q(\Delta \omega )=[s+i(\Delta \omega +\bar{\delta})][s+i(\Delta
\omega +\delta )+\Gamma /2]+\Omega ^{2}$. The frequency integral in equation
(\ref{eq. K}) is integrated over the range $\pm \infty $ for convenience.
The matrices $\mathbf{K}_{r}$, $\mathbf{K}_{i}$ are determined from the
kernel $K$ and the matrices $\mathbf{d}_{r}$, $\mathbf{d}_{i}$ from an
analogous equation for $d$. The amplitude $b_{2}(t)$ is obtained by
numerical integration.

We note that in the special case of a structure resonant with both the
transitions, $\omega _{1}=\omega _{2}=\omega _{0}=\omega _{1}^{c}$ and $%
\delta =\bar{\delta}=0$. The kernel then reduces to the case studied in \cite%
{Dalton03a}, i.e. 
\begin{equation}
K(\Delta \omega ,\Delta \omega ^{\prime })=\frac{\Gamma \Omega ^{2}}{2\pi }%
\frac{(s+i\Delta \omega ++\Gamma /2)(2s+i(\Delta \omega +\Delta \omega
^{\prime })}{s(\Delta \omega ^{\prime }{}^{2}+(\Gamma /2)^{2})(s+i(\Delta
\omega +\Delta \omega ^{\prime }))Q(\Delta \omega )},
\label{eq. K High Q RES}
\end{equation}%
where now $Q(\Delta \omega )=(s+i\Delta \omega )(s+i\Delta \omega +\Gamma
/2)+\Omega ^{2}$.

\begin{figure}
 \includegraphics[width=8cm]{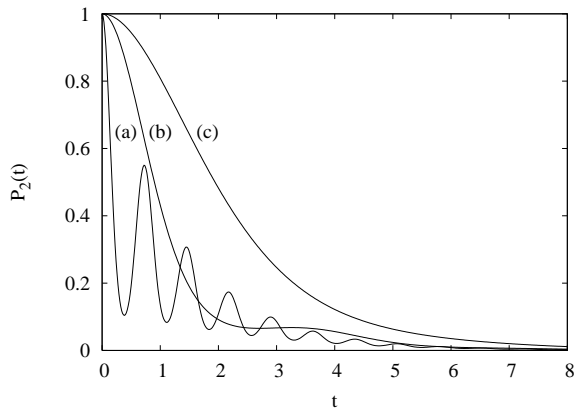}
\caption[Fig2]{ High-Q cavity model. Time evolution of the probability of
finding the system in state 2; $P(t)=|b_{2}(t)|^{2}$. The reservoir
structure function is given by equation (\protect\ref{eq. RSF High Q 2}),
with parameters $\Gamma $ =1 and: (a) $\Omega $ = 5.0; (b) $\Omega $ = 1.0;
and (c) $\Omega $ = 0.5, in scaled units. The grid size for the discretised
kernel was 150$\times 150$ chosen with a range of $\pm $30 for $\Delta 
\protect\omega $ and $\Delta \protect\omega ^{\prime }$ in scaled units. The
case of resonance is shown, where each transition is resonant with the
reservoir structure: $\bar{\protect\delta}=\protect\delta =0$. Results were
computed with the integral equation method of section  \protect\ref%
{sec:num_integral}. }
\label{Fig2}
\end{figure}

Typical results for the time-dependent upper state probability are shown in
figure \ref{Fig2}, where the parameters used are $\Gamma $ =1 and (a) $%
\Omega $ =5 (b) $\Omega $ =1 (c) $\Omega $ =0.5 (effectively in scaled
units). The damped oscillation in case (a) is typical of \emph{non-Markovian
decay}. In case (b) the oscillations weaken. Further reduction in coupling
in case (c) removes the oscillations and the decay is closer to
exponential.

The case just described applies when both transitions are coupled to a \emph{%
single} reservoir. A different situation applies when the two transitions
are coupled to \emph{two separate} reservoirs, and the results for this case
are presented in \cite{Dalton03a}. The equations are then simpler, and the
integral equation for the amplitudes can be solved analytically. The decay
of the excited state is now less oscillatory. Unlike the single reservoir
case, the two photons emitted can now be distinguished.

\begin{figure}
 \includegraphics[width=8cm]{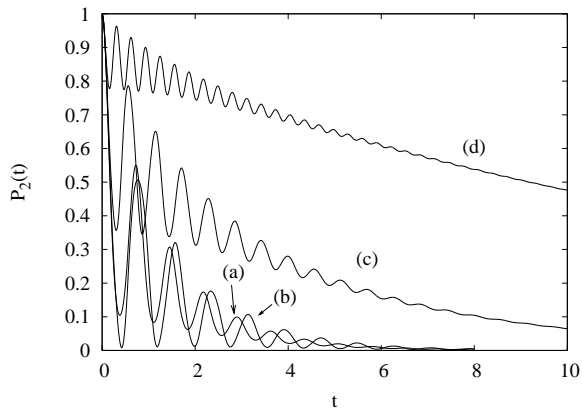}
\caption[FigA2]{ High-Q cavity model. Time evolution of the probability of
finding the system in state 2; $P(t)=|b_{2}(t)|^{2}$. The reservoir
structure function is given by equation (\protect\ref{eq. RSF High Q 2}),
with parameters $\Gamma $ =1 and $\Omega $ = 5.0 in scaled units. The atomic
transitions have equal frequency: $\bar{\protect\delta}=0$. The centre of
the resonance is offset from the atomic frequency by the detunings in
(a)-(d) such that $\protect\delta = 0, 5, 10, 20$. The results for (a) and
(b) are computed using the integral equation method of section  \protect\ref%
{sec:num_integral} and the results for (c) and (d) are computed using the
master equation method discussed in section \protect\ref{sec:ME}. }
\label{FigA2}
\end{figure}

\begin{figure}
 \includegraphics[width=8cm]{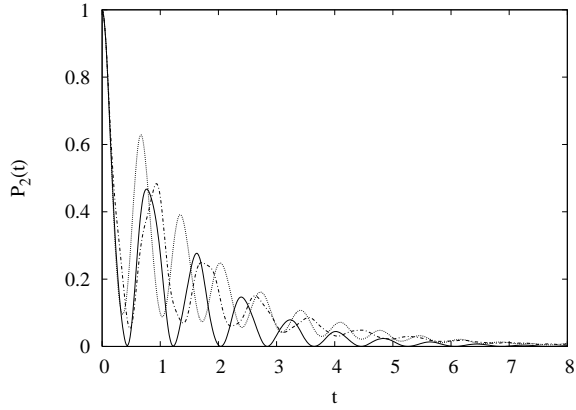}
\caption[FigA3]{ High-Q cavity model. Time evolution of the probability of
finding the system in state 2; $P(t)=|b_{2}(t)|^{2}$. The reservoir
structure function is given by equation (\protect\ref{eq. RSF High Q 2}),
with parameters $\Gamma $ =1 and $\Omega $ = 5.0 in scaled units. The
detuning $\protect\delta =4$ and the off-set is $\bar{\protect\delta}=0$
(solid line), $+\protect\delta$ (chain), and $-\protect\delta$ (dotted). The
results are computed using the integral equation method of section  \protect
\ref{sec:num_integral}. }
\label{FigA3}
\end{figure}

The typical effects of detunings on the upper state probability are
presented in figures \ref{FigA2} and \ref{FigA3}. Figure \ref{FigA2} shows
the case where the two atomic transitions have identical frequency and are
detuned from the reservoir structure by increasing amounts [curves (a)-(d)].
It is seen that the oscillation frequency initially decreases, but then
increases with the detuning. The damping of the oscillations remains about
the same, but the decay of the population itself reduces as the detuning
increases. For the large detuning cases $\delta =10,20$ (curves (c), (d) in
figure \ref{FigA2}) the integral equation method was very slow, and
therefore these curves were computed using the master equation method to be
discussed in Section \ref{sec:ME}. As we will see, this method is
equivalent to the integral equation method since the same reservoir
structure function is involved. Figure \ref{FigA3} shows the situation for a
fixed cavity mode detuning $\delta =4$, but with atomic transition
frequencies that may be different. The chained and dotted curves in figure %
\ref{FigA3} show the effect of unequal atomic transition frequencies. In
fact, since $\bar{\delta}=\pm \delta $ for these two cases, we can see from
equation (\ref{def.deltabar}), and $\delta $ used in equation (\ref{eq. RSF
High Q 2}), that $\bar{\delta}=\pm \delta $ corresponds to the Lorentzian
structure being resonant with the lower atomic transition, or the upper
transition, respectively. The effect on the population dynamics is similar
in both cases---the population is lost slightly less quickly, because of
less efficient coupling, and the frequency of the oscillations changes. The
frequency of population oscillation is reduced when the structure is
resonant with the lower transition, and increased when it is resonant with
the upper transition. Note that for equal atomic transition frequencies both
figure \ref{FigA2} and figure \ref{FigA3} display a case where the
population drops close to zero periodically. This is for $\delta =5,4$
respectively.

\subsection{Case B: Photonic Band-Gap Model}

\label{sec:pbg}

In this section we consider a simple model of a three-level atom coupled to
a \emph{photonic band-gap system}. For simplicity, we again assume equal
coupling constants and transition frequencies. Thus $g_{\lambda
1}=g_{\lambda 2}=g_{\lambda }$, so that $\eta =1$. In this case the rapid
frequency dependence of the reservoir structure function is due to \emph{%
mode frequency gaps}, as a quasimode theory of such systems demonstrates 
\cite{Dalton02a}. A single reservoir structure function given as the
difference of two Lorentzians is assumed \cite{Garraway97a} to approximately
represent the frequency gap, which is assumed to be detuned from the average
atomic transition frequency by $\delta $ (which is the same for both
Lorentzians). In this case we let $Z_{1}=\Omega _{1}^{2}$ and $Z_{2}=-\Omega
_{2}^{2}$ with $\omega _{1}^{c}=\omega _{2}^{c}=\omega ^{c}$, so that $%
\delta _{1}=\delta _{2}=\delta $. Then the reservoir structure function is
given by: 
\begin{eqnarray}
R_{1} &=&R_{2}=R=\rho _{\lambda }g_{\lambda }^{2}  \label{eq. RSF PBG 1} \\*
&=&\frac{\Omega _{1}^{2}}{2\pi }\frac{\Gamma _{1}}{(\omega _{\lambda
}-\omega ^{c})^{2}+(\Gamma _{1}/2)^{2}}-\frac{\Omega _{2}^{2}}{2\pi }\frac{%
\Gamma _{2}}{(\omega _{\lambda }-\omega ^{c})^{2}+(\Gamma _{2}/2)^{2}}, 
\nonumber \\*
&\equiv &\frac{\Omega _{1}^{2}}{2\pi }\frac{\Gamma _{1}}{(\Delta \omega
-\delta )^{2}+(\Gamma _{1}/2)^{2}}-\frac{\Omega _{2}^{2}}{2\pi }\frac{\Gamma
_{2}}{(\Delta \omega -\delta )^{2}+(\Gamma _{2}/2)^{2}}.
\label{eq. RSF PBG 2}
\end{eqnarray}%
The Lorentzians have coupling strengths $\Omega _{1},$ $\Omega _{2}$ and
reservoir structure widths $\Gamma _{1},\,\Gamma _{2}$, where 
\begin{eqnarray}
\frac{\Omega _{1}^{2}}{\Gamma _{1}} &=&\frac{\Omega _{2}^{2}}{\Gamma _{2}}
\label{eq. PBG Condition 1} \\
\Gamma _{2} &<&\Gamma _{1}.  \label{eq. PBG Condition 2}
\end{eqnarray}%
These conditions are required so that $R$ is always positive and has a zero
where $\Delta \omega =\delta $. This localised zero models the photonic
band-gap (see \cite{Garraway97a,Garraway97b}). The atomic transition
frequencies might not be equal to each other, with the difference being
represented by the parameter $\bar{\delta}$, equation (\ref{def.deltabar}),
but if $\bar{\delta}=\delta =0$ then both transition frequencies would be
resonant with the zero in the reservoir structure function equation (\ref%
{eq. RSF PBG 2}).

The kernel $K$ and the function $d$ are now given by (see equations (\ref%
{eq. K},\ref{eq. D})):%
\begin{eqnarray}
K(\Delta \omega ,\Delta \omega ^{\prime }) &=&\frac{1}{2\pi }\frac{\left(
\Omega _{1}^{2}\Gamma _{1}-\Omega _{2}^{2}\Gamma _{2}\right) \left( \Delta
\omega ^{\prime }-\delta \right) ^{2}}{((\Delta \omega ^{\prime }-\delta
)^{2}+(\Gamma _{1}/2)^{2})((\Delta \omega ^{\prime }-\delta )^{2}+(\Gamma
_{2}/2)^{2})}  \nonumber \\
&&\times \frac{(2s+i(\Delta \omega +\Delta \omega ^{\prime }))}{s(s+i(\Delta
\omega +\Delta \omega ^{\prime }))}  \nonumber \\
&&\times \frac{(s+i(\Delta \omega +\delta )+\Gamma _{1}/2)(s+i(\Delta \omega
+\delta )+\Gamma _{2}/2)}{Q_{3}(s+i(\Delta \omega +\delta ))}
\label{eq. K PBG} \\
d(\Delta \omega ) &=&\frac{-i}{s}\frac{(s+i(\Delta \omega +\delta )+\Gamma
_{1}/2)(s+i(\Delta \omega +\delta )+\Gamma _{2}/2)}{Q_{3}(s+i(\Delta \omega
+\delta ))},  \label{eq. D PBG}
\end{eqnarray}

with 
\begin{eqnarray}
Q_{3}(x) &=& [ x + i(\bar{\delta}-\delta)](x + \Gamma _{1}/2)(x + \Gamma
_{2}/2) + x( \Omega _{1}^{2}-\Omega _{2}^{2} ) .  \label{eq. Q PBG}
\end{eqnarray}%
In equations (\ref{eq. K PBG})-(\ref{eq. Q PBG}) we have used relation (\ref%
{eq. PBG Condition 1}).

The matrices $\mathbf{K}_{r}$, $\mathbf{K}_{i}$ are determined from the
kernel $K$ and the matrices $\mathbf{d}_{r}$, $\mathbf{d}_{i}$ from an
analogous equation for $d$. The amplitude $b_{2}(t)$ is obtained by
numerical integration.

\begin{figure}
  \includegraphics[width=8cm]{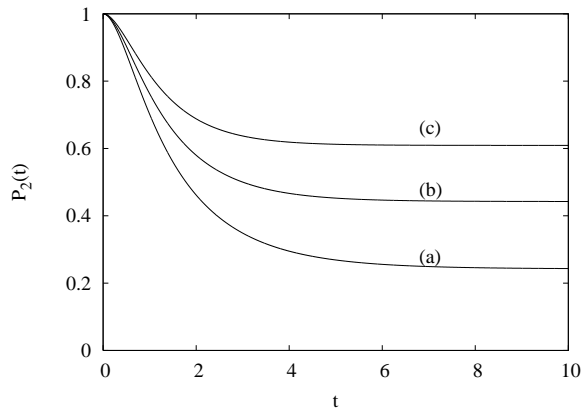}
\caption[B1]{ Photonic band-gap model. Time evolution of the probability of
finding the system in state 2; $P(t)=|b_{2}(t)|^{2}$. The reservoir
structure function is given by equation (\protect\ref{eq. RSF PBG 2}), with
parameters $\Gamma_1 $ =4 and $\Omega_1 =1$ and: (a) $\Gamma_2 $ =1 (b) $%
\Gamma_2 $ =1.5 (c) $\Gamma_2 $ =2.0 in scaled units. The case of resonance
is shown, where each transition is resonant with the gap: $\bar{\protect%
\delta}=\protect\delta=0$. }
\label{FigB1}
\end{figure}

Typical results are shown in figure \ref{FigB1} for the case of two equal
atomic transition frequencies, also coincident with the gap frequency. Thus $%
\omega _{1}=\omega _{2}=\omega ^{c}$, so that $\delta =\widetilde{\delta }=0$%
. The parameters used are $\Gamma _{1}$ $=4$ and $\Omega _{1}=1$ with $%
\Gamma _{2}$ =1.0, 1.5 and 2.0. Non-Markovian decay is seen for small $t$ as
the initial quadratic behaviour. Population trapping effects are found in
all three cases. The amount of trapped population increases as $\Gamma _{2}$
increases.

\begin{figure}
 \includegraphics[width=8cm]{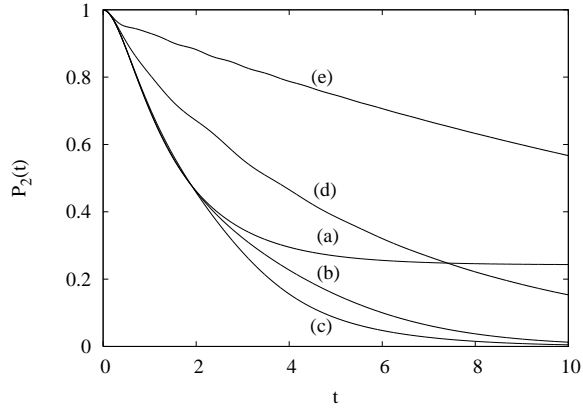}
\caption[B2]{ Photonic band-gap model. Time evolution of the probability of
finding the system in state 2; $P_2(t)=|b_{2}(t)|^{2}$. The reservoir
structure function is given by equation (\protect\ref{eq. RSF PBG 2}), with
parameters $\Gamma_1 $ =4 and $\Omega_1 =1$ and $\Gamma_2 $ =1 in scaled
units. The atomic transitions have equal frequency: $\bar{\protect\delta}=0$%
. The centre of the gap is offset from the atomic frequency by the detunings
in (a)-(e) of $\protect\delta = 0, 0.5, 1, 4$ and $8$. The results for $%
\protect\delta = 0$ to 4 are computed using the integral equation method of
section \protect\ref{sec:num_integral} and the result for $\protect\delta = 8
$ is computed using the master equation method discussed in section \protect
\ref{sec:ME}. }
\label{FigB2}
\end{figure}

\begin{figure}
 \includegraphics[width=8cm]{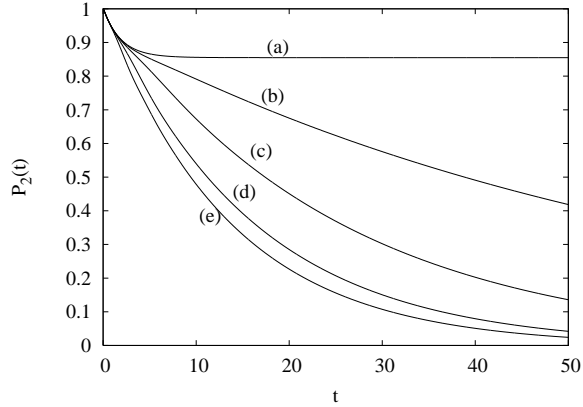}
\caption[B3]{ Photonic band-gap model. Time evolution of the probability of
finding the system in state 2; $P_2(t)=|b_{2}(t)|^{2}$. The reservoir
structure function is given by equation (\protect\ref{eq. RSF PBG 2}), with
parameters $\Gamma_1 $ =50 and $\Omega_1 =1$ and $\Gamma_2 $ =1 in scaled
units. The atomic transitions have equal frequency: $\bar{\protect\delta}=0$%
. The centre of the gap is offset from the atomic frequency by the detunings
in (a)-(e) of $\protect\delta = 0, 0.25, 0.5, 1$ and $2$. These results are
computed using the master equation method discussed in section \protect\ref%
{sec:ME}. }
\label{FigB3}
\end{figure}

Figure \ref{FigB2} shows the effect of detuning the two (equal) atom
transition frequencies from the centre of the gap. The parameters used are
as in figure \ref{FigB1} with $\Gamma _{2}$ =1. Again, for the large
detuning case $\delta =8$ (curve (e) in figure \ref{FigB2}) the results were
calculated using the master equation method of Section \ref{sec:ME}. We see
that there is a loss of the population trapping effect, although for larger
detunings $\delta $ the rate of decay is considerably reduced. 
However, this reduction can be attributed to the decrease in the density of
states when detuned far from the gap and outside the scope of the second,
positive, Lorentzian. In that sense it is an artifact of the model. The
effect is eliminated for the parameters shown in figure \ref{FigB3} where $%
\Gamma _{1}$ is comparatively large. The parameters are as in figure \ref%
{FigB2} except that $\Gamma _{1}=50$ and the detunings selected are
different. All the graphs in figure \ref{FigB2} were computed using the
master equation method. In this case only the loss of the trapping is seen
as the detuning increases, and the curves appear to saturate as the
background density of states flattens out for the larger detunings.

\section{MASTER EQUATION THEORY}

\label{sec:ME}

\subsection{Pseudomode Treatment}

\label{sec:Pseudomode}

We have previously found that Lorentzian reservoir structures can be
connected, in a non-perturbative way, to Lindblad master equations using the
pseudomode approach \cite{Garraway97a,Garraway97b}, which involves
considering the positions and residues of the poles of the reservoir
structure function in the complex frequency plane. Each pole is associated
with a pseudomode. 

This approach was applied to the case of a two-level atom in a high-Q cavity
with a single resonance of width $\Gamma _{1}$ \cite{Garraway97a,Garraway97b}%
. The system takes on the dynamics of a damped Jaynes-Cummings model where
the atom exchanges energy with a pseudomode, from which the energy is lost.
In the case of our three-level atom in a high-Q cavity the pseudomode
approach would result in a master equation of the form 
\begin{equation}
\frac{\mbox{d}}{\mbox{d}t}\hat{\rho}=-\frac{i}{\hbar }\left[ \widehat{H}%
_{0}^{CAV},\hat{\rho}\right] -\frac{\Gamma _{1}}{2}\left( \hat{a}%
_{1}^{\dagger }\hat{a}_{1}\hat{\rho}-2\hat{a}_{1}\hat{\rho}\hat{a}%
_{1}^{\dagger }+\hat{\rho}\hat{a}_{1}^{\dagger }\hat{a}_{1}\right) ,
\label{Eq. PseudoMEHighQ}
\end{equation}%
with the Hamiltonian 
\begin{equation}
\widehat{H}_{0}^{CAV}=\hbar \left[ \omega _{1}|1\rangle \langle 1|+(\omega
_{1}+\omega _{2})|2\rangle \langle 2|+\omega ^{c}\hat{a}_{1}^{\dagger }\hat{a%
}_{1}+\widehat{H}_{AF}^{(1)}\right] \,.  \label{ho:dampedjcm_1mode}
\end{equation}%
Here the term $\widehat{H}_{AF}^{(1)}$ represents the atom-pseudomode
coupling 
\begin{equation}
\widehat{H}_{AF}^{(1)}=\hbar \Omega \left( \hat{a}_{1}^{\dagger }|0\rangle
\langle 1|+\hat{a}_{1}|1\rangle \langle 0|+\hat{a}_{1}^{\dagger }|1\rangle
\langle 2|+\hat{a}_{1}|2\rangle \langle 1|\right) \;.  \label{ho:AF1}
\end{equation}%
Comparison of the calculated results shows that this master equation
reproduces the results of Section \ref{sec:hiq} for the atomic populations
obtained by the integral equation method. 

The pseudomode approach was also applied to the case of a two-level atom in
a photonic band-gap system \cite{Garraway97a}. In the case of cascade atom
in the band-gap material, as in Section \ref{sec:pbg}, this approach \cite%
{Garraway97a,Garraway97b} would, in general, result in a two-mode master
equation of the form: 
\begin{eqnarray}
\frac{\mbox{d}}{\mbox{d}t}\hat{\rho} &=&-\frac{i}{\hbar }\left[ \widehat{H}%
_{0}^{BG},\hat{\rho}\right] -\frac{\Gamma _{1}^{\prime }}{2}\left( \hat{a}%
_{1}^{\dagger }\hat{a}_{1}\hat{\rho}-2\hat{a}_{1}\hat{\rho}\hat{a}%
_{1}^{\dagger }+\hat{\rho}\hat{a}_{1}^{\dagger }\hat{a}_{1}\right)  \nonumber
\\
&&-\frac{\Gamma _{2}^{\prime }}{2}\left( \hat{a}_{2}^{\dagger }\hat{a}_{2}%
\hat{\rho}-2\hat{a}_{2}\hat{\rho}\hat{a}_{2}^{\dagger }+\hat{\rho}\hat{a}%
_{2}^{\dagger }\hat{a}_{2}\right)  \label{dampedjcmbgm_2mode}
\end{eqnarray}%
with the Hamiltonian 
\begin{eqnarray}
\widehat{H}_{0}^{BG} &=&\hbar \left[ \omega _{1}|1\rangle \langle 1|+(\omega
_{1}+\omega _{2})|2\rangle \langle 2|+\omega ^{c}\hat{a}_{1}^{\dagger }\hat{a%
}_{1}+\omega ^{c}\hat{a}_{2}^{\dagger }\hat{a}_{2}\right]  \nonumber \\*
&&+\hbar \frac{\sqrt{\Gamma _{1}\Gamma _{2}}}{2}\left( \hat{a}_{1}^{\dagger }%
\hat{a}_{2}+\hat{a}_{1}\hat{a}_{2}^{\dagger }\right)  \nonumber \\*
&&+\widehat{H}_{AF}^{(2)}.  \label{ho:dampedjcm_2mode}
\end{eqnarray}%
This time the atom-pseudomode coupling is to the pseudomode labelled `2',
i.e. 
\begin{equation}
\widehat{H}_{AF}^{(2)}=\hbar \Omega _{PBG}\left( \hat{a}_{2}^{\dagger
}|0\rangle \langle 1|+\hat{a}_{2}|1\rangle \langle 0|+\hat{a}_{2}^{\dagger
}|1\rangle \langle 2|+\hat{a}_{2}|2\rangle \langle 1|\right) \;,
\label{ho:AF2}
\end{equation}%
similarly to equation (\ref{ho:AF1}). The two modes (or pseudo modes) of
this case are coupled to each other in the Hamiltonian (\ref%
{ho:dampedjcm_2mode}). For the case when the reservoir structure function
has a zero, as has been prescribed by equation (\ref{eq. PBG Condition 1}),
we find that $\Gamma _{1}^{\prime }=0$. However, $\Gamma _{2}^{\prime }$ is
non-zero and is related to the large $\omega $ behaviour of the reservoir
structure function $R$. This parameter, and the atom-pseudomode coupling
strength are given by 
\begin{eqnarray}
\Gamma _{2}^{\prime } &=&\Gamma _{1}+\Gamma _{2}\;, \\
\Omega _{PBG} &=&\sqrt{\Omega _{1}^{2}-\Omega _{2}^{2}}  \label{eq:omPBG}
\end{eqnarray}%
which are both positive. Again, comparison of the calculated results shows
that the master equation (\ref{dampedjcmbgm_2mode}) reproduces the results
of Section \ref{sec:pbg} for the atomic populations obtained by the integral
equation method. For the full density matrix, 36 coupled differential
equations are required with the two excitations produced from the
three-level ladder system.

\subsection{Quasimode Treatment}

\label{sec:Quasimode}

Markovian master equations describing Lorentzian reservoir structures can
also be obtained based on quasimodes \cite{Dalton01a,Dalton01b}, which are
identified with approximate modes of the EM field. The basic concepts and
equations for quasimode theory are set out in the Appendix. For the present
paper we consider the case where the continuum quasimode density $\varrho
_{c}(\Delta )$ and the $n$ discrete-continuum coupling constants $%
W_{i}(\Delta )$ are slowly varying functions of the continuum quasimode
frequency $\Delta $. In this situation we obtain Markovian master equations
(equation (\ref{Eq. MarkovMaster})) for the density operator $\hat{\rho}$
describing the system consisting of the atom plus the $n$ discrete
quasimodes. The continuum quasimodes constitute the reservoir. In addition
simple expressions are obtained for the reservoir structure function $%
R_{k}(\omega )$ for the true modes [equation (\ref{Eq. ResSpectDens})].
These results will now be applied to the high-Q cavity and photonic band-gap
cases. 

For the high-Q cavity case, only a single discrete quasimode is involved.
The master equation for the cascade atom plus single discrete quasimode
system is then 
\begin{equation}
\frac{d}{dt}\hat{\rho}=\frac{-i}{\hbar }\,[\widehat{H}_{S},\hat{\rho}]+\frac{%
\Gamma }{2}\,\{[\hat{a}_{1},\hat{\rho}\hat{a}_{1}^{\dagger }]+[\hat{a}_{1}%
\hat{\rho},\hat{a}_{1}^{\dagger }]\},  \label{Eq. HighQME}
\end{equation}%
where the decay rate $\Gamma $ is given by%
\begin{equation}
\Gamma =2\pi \varrho _{c}\,\left\vert W_{1}\right\vert ^{2},
\label{Eq. DecayRate}
\end{equation}%
and the system Hamiltonian is%
\begin{eqnarray}
\widehat{H}_{S} &=&\sum_{k=1,2}\eta _{k}\,\hbar \omega _{k}\,(\widehat{%
\sigma }_{k}^{+}\widehat{\sigma }_{k}^{-}-\widehat{\sigma }_{k}^{-}\widehat{%
\sigma }_{k}^{+})+\,\hbar \nu _{1}\,\hat{a}_{1}^{\dagger }\,\hat{a}_{1} 
\nonumber \\
&&+\sum_{k=1,2}\,(\hbar \lambda _{k1}^{\ast }\hat{a}_{1}\,\widehat{\sigma }%
_{k}^{+}+H.c),  \label{Eq. HighQSysHam}
\end{eqnarray}%
with $\widehat{\sigma }_{1}^{+}=|1\rangle \langle 0|$, $\widehat{\sigma }%
_{2}^{+}=|2\rangle \langle 1|$, $\eta _{1}=\frac{2}{3}+\frac{1}{3}\frac{%
\omega _{2}}{\omega _{1}}$, $\eta _{2}=\frac{2}{3}+\frac{1}{3}\frac{\omega
_{1}}{\omega _{2}}$. The atomic Hamiltonian is that given in
equation (\ref{eq:hamil}), apart from an additive constant $(2\hbar \omega
_{1}+\hbar \omega _{2})/3$. The choice $\lambda _{11}=\lambda _{21}=\Omega $
(real) is natural since we are dealing with a situation where both
transitions have equal coupling constants with the EM field, and we choose $%
\nu _{1}=\omega ^{c}$ to have the quasimode frequency coincide with the peak
in the reservoir structure function. These choices then result in the same
master equation as equation (\ref{Eq. PseudoMEHighQ}). A straightforward
evaluation of the reservoir structure function from equation (\ref{Eq.
ResSpectDens}) gives%
\begin{equation}
R_{k}(\omega )=\Omega ^{2}\,\frac{\Gamma }{2\pi }\,\frac{1}{(\omega -\omega
^{c})^{2}+(\Gamma /2)^{2}}  \label{Eq. HighQResStrFn}
\end{equation}%
for both transitions. This result is the same as in equations (\ref{eq. RSF
High Q 1},\ref{eq. RSF High Q 2}). Thus the high-Q cavity case can be
treated via quasimode theory involving a single discrete quasimode, which
may be identified as the cavity mode.

For the photonic band-gap case, two discrete quasimodes are involved. There
is now more choice for the parameters in the quasimode model and it turns
out we can choose these to obtain the same reservoir structure function as
in equations (\ref{eq. RSF PBG 1},\ref{eq. RSF PBG 2}) as well as generating
a master equation which is the same as equation (\ref{dampedjcmbgm_2mode})
obtained from pseudomode theory. However, to achieve this we must first
introduce two new discrete quasimodes, whose annihilation operators $%
\widehat{b}_{1}$, $\widehat{b}_{2}$ are defined by%
\begin{eqnarray}
\widehat{b}_{1} &=&\kappa ^{-1/2}\,\pi \rho _{c}\,(-W_{2}\,\hat{a}%
_{1}+W_{1}\,\hat{a}_{2})  \label{Eq. NewQM1} \\
\widehat{b}_{2} &=&\kappa ^{-1/2}\,\pi \rho _{c}\,(W_{1}^{\ast }\,\hat{a}%
_{1}+W_{2}^{\ast }\,\hat{a}_{2})  \label{Eq. NewQM2} \\
\kappa &=&\frac{1}{2}(\Gamma _{1}+\Gamma _{2}),
\end{eqnarray}%
where the decay rates are%
\begin{eqnarray}
\Gamma _{1} &=&2\pi \varrho _{c}\,\left\vert W_{1}\right\vert ^{2}
\label{Eq. DecayRate1} \\
\Gamma _{2} &=&2\pi \varrho _{c}\,\left\vert W_{2}\right\vert ^{2}.
\label{Eq. DecayRate2}
\end{eqnarray}%
Note that the $\widehat{b}_{1}$, $\widehat{b}_{2}$ and their adjoints
satisfy the standard Bose commutation rules, which is a consequence of the
matrix involved in (\ref{Eq. NewQM1}, \ref{Eq. NewQM2}) being
unitary.

From equation (\ref{Eq. MarkovMaster}) we immediately see that the master
equation is given by%
\begin{equation}
\frac{d}{dt}\hat{\rho}=\frac{-i}{\hbar }\,[\widehat{H}_{S},\hat{\rho}]+\frac{%
1}{2}(\Gamma _{1}+\Gamma _{2})\{[\widehat{b}_{2},\hat{\rho}\widehat{b}%
_{2}^{\dagger }]+[\widehat{b}_{2}\hat{\rho},\widehat{b}_{2}^{\dagger }]\},
\label{Eq. PBGME}
\end{equation}%
showing that relaxation only involves the new quasimode with annihilation
operator $\widehat{b}_{2}$, the decay rate being the sum of the original
rates $\Gamma _{1}$ and $\Gamma _{2}$. This result for the relaxation terms
is the same as in the pseudomode master equation (\ref{dampedjcmbgm_2mode}).

The system Hamiltonian $\widehat{H}_{S}$ can then be written in terms of the
new quasimode operators by inverting equations (\ref{Eq. NewQM1}, \ref{Eq.
NewQM2}) to give $\hat{a}_{1}$, $\hat{a}_{2}$ in terms of $\widehat{b}_{1}$, 
$\widehat{b}_{2}$ and substituting in equation (\ref{Eq. SystemHam}). There
are still many choices that can be made for the quasimode frequencies $\nu
_{i}$ $(i=1,2)$, the magnitude and phases of the atom-quasimode coupling
constants $\lambda _{ki}$ $(k=1,2;i=1,2)$, the quasimode-quasimode coupling
constant $V_{12}$ and the phase for discrete-continuum coupling constants $%
W_{i}$ $(i=1,2)$. The magnitude of the $W_{i}$ are equivalent to the decay
rates $\Gamma _{i}$. The atomic term in the system Hamiltonian $\widehat{H}%
_{S}$ is unaltered%
\begin{equation}
\widehat{H}_{S1}=\sum_{k=1,2}\eta _{k}\,\hbar \omega _{k}\,(\widehat{\sigma }%
_{k}^{+}\widehat{\sigma }_{k}^{-}-\widehat{\sigma }_{k}^{-}\widehat{\sigma }%
_{k}^{+}).  \label{Eq. PBGAtomHam}
\end{equation}%

Since the photonic band-gap model involves two Lorentzians with the same
centre frequency $\omega ^{c}$ it is natural to choose the discrete
quasimode frequencies $\nu _{i}$ to be both equal to $\omega ^{c}$, and
since both atomic transitions have equal coupling constants with the EM\
field it is natural to choose the atom-quasimode coupling constants to be
independent of the transition. Thus
\begin{eqnarray}
\nu _{1} &=&\nu _{2}=\omega ^{c}  \label{Eq. PBGCond1} \\
\lambda _{k1} &=&\lambda _{1},\quad \lambda _{k2}=\lambda _{2}.
\label{Eq. PBGCond2}
\end{eqnarray}%
We also introduce the various phase factors via the definitions%
\begin{eqnarray}
V_{12} &=&\left\vert V_{12}\right\vert \,\exp (i\xi _{12})
\label{Eq. PBG PhaseDefn1} \\
W_{1} &=&\left\vert W_{1}\right\vert \,\exp (i\phi _{1}), \quad
W_{2}=\left\vert W_{2}\right\vert \,\exp (i\phi _{2})
\label{Eq. PBG PhaseDefn2} \\
\lambda _{1} &=&\left\vert \lambda _{1}\right\vert \,\exp (i\theta
_{1}),\quad\quad \lambda _{2}=\left\vert \lambda _{2}\right\vert \,\exp
(i\theta _{2}).  \label{Eq. PBG PhaseDefn3}
\end{eqnarray}%

In view of (\ref{Eq. PBGCond1}) and the unitary relation between the new and
original discrete quasimode operators, it follows that the quasimode energy
term in $\widehat{H}_{S}$ is given by 
\begin{equation}
\widehat{H}_{S2}=\hbar \omega ^{c}\,(\widehat{b}_{1}^{\dagger }\,\widehat{b}%
_{1}+\widehat{b}_{2}^{\dagger }\,\widehat{b}_{2})\mathbf{.}
\label{Eq. PBGQMHam}
\end{equation}%

To obtain the required atom-quasimode interaction we choose 
\begin{eqnarray}
\left\vert V_{12}\right\vert &=&\frac{1}{2}\sqrt{\Gamma _{1}\Gamma _{2}}
\label{Eq. PBGCond3} \\
\xi _{12} &=&\phi _{1}-\phi _{2}+(2\nu +1)\frac{\pi }{2}\qquad (\nu =0,\pm
1,\pm 2,..)  \label{Eq. PBGCond4} \\
\phi _{1}+\phi _{2} &=&(2\nu +1)\frac{\pi }{2}-\mu \,2\pi \qquad (\mu =0,\pm
1,\pm 2,..),  \label{Eq. PBGCond5}
\end{eqnarray}%
and find that the quasimode-quasimode interaction is given by 
\begin{equation}
\widehat{H}_{S3}=\hbar \,\frac{1}{2}\sqrt{\Gamma _{1}\Gamma _{2}}\,(\widehat{%
b}_{1}^{\dagger }\,\widehat{b}_{2}+\widehat{b}_{2}^{\dagger }\,\widehat{b}%
_{1})\mathbf{.\ }  \label{Eq. PBGQMQMIntn}
\end{equation}%

To obtain the required atom-quasimode coupling together with condition (\ref%
{Eq. PBGCond2}) we choose 
\begin{eqnarray}
\left\vert \lambda _{1}\right\vert &=&\Omega _{1}\sqrt{\frac{\Omega
_{1}^{2}-\Omega _{2}^{2}}{\Omega _{1}^{2}+\Omega _{2}^{2}}},\qquad
\left\vert \lambda _{2}\right\vert =\Omega _{2}\sqrt{\frac{\Omega
_{1}^{2}-\Omega _{2}^{2}}{\Omega _{1}^{2}+\Omega _{2}^{2}}}
\label{Eq. PBGCond6} \\
\theta _{1}-\phi _{1} &=&\theta _{2}-\phi _{2}  \label{Eq. PBGCond7} \\
&=&\xi \,2\pi \qquad (\xi =0,\pm 1,\pm 2,..),  \label{Eq. PBGCond8}
\end{eqnarray}%
where the (real) quantities $\Omega _{1}$, $\Omega _{2}$, $\Gamma _{1}$and $%
\Gamma _{2}$ are related as in equations (\ref{eq. PBG Condition 1}, \ref%
{eq. PBG Condition 2}). 
We thus find that 
\begin{equation}
\widehat{H}_{S4}=\hbar \,\Omega _{PBG}\,\left( \widehat{b}_{2}\,(\widehat{%
\sigma }_{1}^{+}+\widehat{\sigma }_{2}^{+})+\widehat{b}_{2}^{\dagger }\,(%
\widehat{\sigma }_{1}^{-}+\widehat{\sigma }_{2}^{-})\right)
\label{Eq. PBGAtomQMIntn}
\end{equation}%
where, as in equation (\ref{eq:omPBG}), 
\begin{equation}
\Omega _{PBG}=\sqrt{\Omega _{1}^{2}-\Omega _{2}^{2}}.  \label{Eq. PBGOmega}
\end{equation}%

Overall, with $\widehat{H}_{S}=\widehat{H}_{S1}+\widehat{H}_{S2}+\widehat{H}%
_{S3}+\widehat{H}_{S4}$ the system Hamiltonian (and hence the master
equation) is the same as that obtained via the pseudomode approach. The key
feature is that we started with two original coupled discrete quasimodes
with equal frequencies, whose parameters satisfy the conditions in equations
(\ref{Eq. PBGCond1}, \ref{eq. PBG Condition 2}, \ref{Eq. PBGCond3}, \ref{Eq.
PBGCond4}, \ref{Eq. PBGCond5}, \ref{Eq. PBGCond6}, \ref{Eq. PBGCond7}, \ref%
{Eq. PBGCond8}). Two new discrete quasimodes were introduced, also coupled
and with equal frequencies. However, the first of these new quasimodes $%
\widehat{b}_{1}$ is neither coupled to the atomic transitions nor involved
in relaxation. The other new quasimode $\widehat{b}_{2}$ is coupled to the
atomic transitions and is involved in relaxation with a rate equal to the
sum of the original rates.

Finally, with the conditions referred to in the last paragraph applying, it
is a straightforward process to evaluate the reservoir structure function
from equation (\ref{Eq. ResSpectDens}). We find that this is the same for
both transitions and is given by%
\begin{eqnarray}
R_{k}(\omega ) &=&R(\omega )  \nonumber \\
&=&\Omega _{1}^{2}\,\frac{\Gamma _{1}}{2\pi }\,\frac{1}{(\omega -\omega
^{c})^{2}+(\Gamma _{1}/2)^{2}}  \nonumber \\
&&-\Omega _{2}^{2}\,\frac{\Gamma _{2}}{2\pi }\,\frac{1}{(\omega -\omega
^{c})^{2}+(\Gamma _{2}/2)^{2}}.  \label{Eq. PBGResSpectDens2}
\end{eqnarray}%
Thus the photonic band-gap case can be treated via quasimode theory
involving two coupled discrete quasimodes of equal frequency, where one may
be identified as a band-gap mode and the other as a background mode.

\section{SUMMARY}

\label{sec:Summary}

The dynamical behaviour of a three level cascade atom coupled to a
structured reservoir (typically of EM field modes), and initially in the
upper state has been analysed via Laplace transform and exact master
equation methods. In the Laplace transform approach, the atomic density
operator is determined from the solutions of integral equations, in which
the properties of the structured reservoir only appears via reservoir
structure functions, all essentially given by the product of the mode
density times the square of coupling constants. In the cascade system two
distinct reservoir structure functions are involved. The dependence of the
dynamics solely on reservoir structure functions is required for treating
structured reservoir problems via pseudomode theory, so our results suggest
that it may be possible to extend pseudomode theory to problems involving
more than a single photon excitation of the reservoir.

Solutions of the integral equation involve discretising the frequency space
into $N$ points, so the matrix inversion step involves a matrix with $4N^{2}$
elements compared to, say, $\mathcal{O}(N^{4})$ elements represented by the
original coupled amplitude equations in an equivalent discretised form.
Considerations based on analytic continuation were required for obtaining
solutions of the problem using numerical methods, including carrying out an
inverse Laplace transform. For some oscillatory problems (specifically, with
large detunings) there were numerical difficulties with the integral
equation approach. 

Two important physical situations have been studied using these numerical
methods. The first applies to a simple model of a three-level atom with
identical cascade transitions coupled to a single high-Q cavity mode,
involving a single Lorentzian reservoir structure function. Non-Markovian
oscillatory decay of the excited state has been demonstrated. The second
applies to a model of a three-level atom with identical cascade transitions
coupled to a photonic band-gap system, involving a single reservoir
structure function modelled simply as the difference between two Lorentzian
functions. Again we have shown the non-Markovian oscillatory decay of the
excited state and population trapping effects.

We have also studied both the high-Q cavity and the photonic band-gap cases
via the master equation method. The master equation can be obtained both
from the pseudomode approach and the quasimode approach, and numerical
calculations of the upper state probability via the master equation method
give the same results as the integral equation method based on essential
states. This shows that complicated non-Markovian decays into structured EM
field reservoirs can be described by Markovian models in which the atomic
system is augmented by one or two pseudomodes or quasimodes, which in the
quasimode approach themselves undergo Markovian relaxation into a flat
reservoir. For the high-Q cavity case the single pseudomode or quasimode may
be identified as the cavity mode, for the photonic band-gap case the two
coupled pseudomodes or quasimodes may be identified as band-gap and
background modes.

\section*{ACKNOWLEDGEMENTS}

This paper is dedicated to the memory of Edwin Power, whose pioneering work on
Quantum Electrodynamics applied to atomic and molecular systems was central
to the development of Quantum Optics. Helpful discussions with R.~Blatt,
A.~Caldeira, H.~Carmichael, M.~Collett and C.~Savage are gratefully
acknowledged.

\appendix

\section*{APPENDIX : QUASIMODE THEORY}

\label{sec: Appendix}

In quasimode theory \cite{Dalton01a,Dalton01b} an atomic system with upward,
downward transition operators $\widehat{\sigma }_{k}^{+}$, $\widehat{\sigma }%
_{k}^{-}$ and transition frequencies $\omega _{k}$ is coupled to a set of $n$
discrete bosonic quasimodes with annihilation, creation operators $\hat{a}%
_{i}$, $\hat{a}_{i}^{\dagger }$ and frequencies $\nu _{i}$. The coupling
constants are denoted $\lambda _{ki}$. The discrete quasimodes may be
coupled to each other with coupling constants $V_{ij}$. In addition, there
is a coupling between the discrete quasimodes and a continuum of quasimodes
whose annihilation, creation operators are $\widehat{b}(\Delta )$, $\widehat{%
b}(\Delta )^{\dag }$ and whose frequencies are $\Delta $. All coupling terms
are in the RWA. The continuum quasimodes have mode density $\varrho
_{c}(\Delta )$ and the coupling constant with the discrete quasimodes are
denoted $W_{i}(\Delta )$. The continuum quasimodes are neither coupled to
each other nor to the atomic system. In Refs.\ \cite{Dalton01a,Dalton01b} the
quasimodes were taken to be approximations to the true EM field modes. The
overall Hamiltonian is given by%
\begin{equation}
\widehat{H}=\widehat{H}_{A}+\widehat{H}_{Q}+\widehat{H}_{AQ},
\label{Eq.QuasiHam}
\end{equation}%
with%
\begin{eqnarray}
\widehat{H}_{A} &=&\sum_{k}\eta _{k}\,\hbar \omega _{k}\,(\widehat{\sigma }%
_{k}^{+}\widehat{\sigma }_{k}^{-}-\widehat{\sigma }_{k}^{-}\widehat{\sigma }%
_{k}^{+})  \label{Eq.AtomHam} \\
\widehat{H}_{Q} &=&\sum_{i}\,\hbar \nu _{i}\,\hat{a}_{i}^{\dagger }\,\hat{a}%
_{i}+\sum_{i\neq j}\,\hbar V_{ij}\,\hat{a}_{i}^{\dagger }\,\hat{a}_{j} 
\nonumber \\
&&+\sum_{i}\int d\Delta \,\varrho _{c}(\Delta )[\hbar W_{i}(\Delta )\,\hat{a}%
_{i}^{\dagger }\mathbf{\ }\widehat{b}(\Delta )+H.c.]  \nonumber \\
&&+\int d\Delta \,\varrho _{c}(\Delta )\,\hbar \Delta \,\widehat{b}(\Delta
)^{\dag }\,\widehat{b}(\Delta )  \label{Eq.QuasiModesHam} \\
\widehat{H}_{AQ} &=&\sum_{k}\sum_{i}\,(\hbar \lambda _{ki}^{\ast }\hat{a}%
_{i}\,\widehat{\sigma }_{k}^{+}+H.c).  \label{Eq.AtomQuasiCoupling}
\end{eqnarray}%
In equation (\ref{Eq.AtomHam}) the quantities $\eta _{k}$ are chosen so that 
$\widehat{H}_{A}$ equals the atomic Hamiltonian given in equation (\ref%
{eq:hamil}), apart from an additive constant $(2\hbar \omega _{1}+\hbar
\omega _{2})/3$. The annihilation and creation operators for the discrete
quasimodes satisfy the Kronecker delta commutation rules, whilst those for
the continuum quasimodes satisfy Dirac delta function commutation rules:%
\begin{eqnarray}
\lbrack \hat{a}_{i},\hat{a}_{j}^{\dagger }] &=&\delta _{ij}  \nonumber \\
\lbrack \widehat{b}(\Delta ),\widehat{b}(\Delta ^{\prime })^{\dag }]
&=&\delta (\Delta -\Delta ^{\prime })/\varrho _{c}(\Delta ).
\label{Eq. CommRules}
\end{eqnarray}%
The $\varrho _{c}$ factor on the right side of the last equation gives
annihilation and creation operators that are dimensionless. 

In Refs.\ \cite{Dalton01a,Dalton01b} the true EM field modes were described by
annihilation and creation operators $\widehat{A}(\omega )$ and $\,\widehat{A%
}(\omega )^{\dag }$ that were obtained by Fano diagonalisation methods. This
involved writing $\widehat{A}(\omega ) $ as a linear combination of the
discrete quasimode annihilation operators $\hat{a}_{i}$ plus an integral of
the continuum annihilation operators $\widehat{b}(\Delta )$, and requiring
that the true mode annihilation operator is an eigenoperator for the
quasimode Hamiltonian (\ref{Eq.QuasiModesHam}) with energy $\hbar \omega $.
Expressions for the atom-quasimode coupling (\ref{Eq.AtomQuasiCoupling})
could then be written in terms of the true mode annihilation and creation
operators (equations (18), (19) of Ref. \cite{Dalton01b}) and a result for
the reservoir structure function $R_{k}(\omega )$ associated with the $k$
atomic transition obtained (equation (50) of Ref. \cite{Dalton01b}). These
results applied to the general case where the continuum quasimode density $%
\varrho _{c}(\Delta )$ and the discrete-continuum coupling constants $%
W_{i}(\Delta )$ were not necessarily slowly varying functions of $\Delta $%
.

For the present paper we are interested in the situation where the dynamics
of the system consisting of the atom plus the $n$ discrete quasimodes is
Markovian. The reservoir consists of the continuum quasimodes. Markovian
behaviour occurs when $\varrho _{c}(\Delta )$ and the $W_{i}(\Delta )$ are
slowly varying functions of $\Delta $ and can be treated as constant. The
flatness of the continuum quasimode density and the coupling constants
results in the reservoir correlation time being short enough for Markovian
behaviour to occur. The density operator $\hat{\rho}$ for the system
consisting of the atom plus the $n$ discrete quasimodes can then be shown to
satisfy the Markovian master equation 
\begin{equation}
\frac{d}{dt}\hat{\rho}=\frac{-i}{\hbar }\,[\widehat{H}_{S},\hat{\rho}%
]+\sum_{ij}\,\pi \varrho _{c}\,W_{i}\,W_{j}^{\ast }\,\{[\hat{a}_{j},\hat{\rho%
}\hat{a}_{i}^{\dagger }]+[\hat{a}_{j}\hat{\rho},\hat{a}_{i}^{\dagger }]\},
\label{Eq. MarkovMaster}
\end{equation}%
where the system Hamiltonian is%
\begin{eqnarray}
\widehat{H}_{S} &=&\sum_{k}\eta _{k}\,\hbar \omega _{k}\,(\widehat{\sigma }%
_{k}^{+}\widehat{\sigma }_{k}^{-}-\widehat{\sigma }_{k}^{-}\widehat{\sigma }%
_{k}^{+})+\sum_{i}\,\hbar \nu _{i}\,\hat{a}_{i}^{\dagger }\,\hat{a}_{i} 
\nonumber \\
&&+\sum_{i\neq j}\,\hbar V_{ij}\,\hat{a}_{i}^{\dagger }\,\hat{a}%
_{j}+\sum_{k}\sum_{i}\,(\hbar \lambda _{ki}^{\ast }\hat{a}_{i}\,\widehat{%
\sigma }_{k}^{+}+H.c).  \label{Eq. SystemHam}
\end{eqnarray}%
\bigskip

In this case of a flat continuum the reservoir structure function is given
by 
\begin{equation}
R_{k}(\omega )=\varrho _{c}\,\frac{\left\vert Q_{n-1}^{k}(\omega
)\right\vert ^{2}}{\left\vert P_{n}(\omega )\right\vert ^{2}},
\label{Eq. ResSpectDens}
\end{equation}%
where%
\begin{eqnarray}
Q_{n-1}^{k}(\omega ) &=&\sum_{ij}\,\lambda _{ki}\,(\omega \mathbf{E}_{n}-%
\mathbf{\Omega })_{ij}^{ADJ}\,W_{j}^{\ast }  \label{Eq. Qn-1} \\
P_{n}(\omega ) &=&\left\vert \omega \mathbf{E}_{n}-\mathbf{\Omega }%
\right\vert -i\pi \varrho _{c}\,\sum_{ij}\,W_{i}\,(\omega \mathbf{E}_{n}-%
\mathbf{\Omega })_{ij}^{ADJ}\,W_{j}^{\ast },  \label{Eq. Pn}
\end{eqnarray}%
and the $n\times n$ matrix $\mathbf{\Omega }$ has elements%
\begin{equation}
\mathbf{\Omega }_{ij}=\nu _{i}\,\delta _{ij}+(1-\delta _{ij})\,V_{ji},
\label{Eq. Omega}
\end{equation}%
with the adjugate matrix $(\omega \mathbf{E}_{n}-\mathbf{\Omega })^{ADJ}$
related to the inverse via%
\begin{equation}
(\omega \mathbf{E}_{n}-\mathbf{\Omega })^{-1}=\frac{(\omega \mathbf{E}_{n}-%
\mathbf{\Omega })^{ADJ}}{\left\vert \omega \mathbf{E}_{n}-\mathbf{\Omega }%
\right\vert }.  \label{Eq.Adjugate}
\end{equation}%
In obtaining this result we note that the frequency shift matrix $%
F_{ij}(\omega )$ (equation (27) of Ref. \cite{Dalton01b}) is zero. 




\begin{thebibliography}{99}
\bibitem{Scully97a} M. O. Scully and M. S. Zubairy, \textit{Quantum Optics},
Cambridge University Press, (Cambridge, 1997)

\bibitem{Auletta01a} G. Auletta, \textit{Foundations and Interpretation of
Quantum Mechanics}, World Scientific, (Singapore, 2001)

\bibitem{Zurek03a} W. H. Zurek, Rev. Mod. Phys. \textbf{75}, 715 (2003)

\bibitem{Zurek93a} W. H. Zurek, S. Habib and J. P. Paz, Phys. Rev. Lett. 
\textbf{70}, 1187 (1993)

\bibitem{Pitaevskii03a} L. Pitaevskii and S. Stringari, \textit{%
Bose-Einstein Condensation}, Clarendon Press, (Oxford, 2003)

\bibitem{Barnett97a} S. M. Barnett and P. M. Radmore, \textit{Methods in
Theoretical Quantum Optics}, (Oxford University Press, Oxford 1997)

\bibitem{Hope00a} J. J. Hope, G. M. Moy, M. J. Collett and C. M. Savage,
Phys. Rev. A \textbf{61}, 023603 (2000)

\bibitem{Strunz04a} W. T. Strunz and T. Yu, Phys. Rev. A \textbf{69}, 052115
(2004)

\bibitem{Breuer01a} H-P. Breuer and B. Kappler, Ann.Phys. (NY) \textbf{291},
36 (2001)

\bibitem{Wiseman03a} P. Warszawski and H. M. Wiseman, J. Opt. B: Quantum
Semiclass. Opt. \textbf{5}, 1 (2003)

\bibitem{Privman04a} V. Privman and D. Tolkunov, e-print cond-matt/0405694

\bibitem{DiVincenzo03a} 
D. P. DiVincenzo and D. Loss, Phys. Rev. B \textbf{71}, 035318 (2005)

\bibitem{Dalton03b} B. J. Dalton, J. Mod. Opt. \textbf{50}, 951 (2003)

\bibitem{Braun01a} D. Braun, F. Haake and W. T. Strunz, Phys. Rev. Lett. 
\textbf{86}, 2913 (2001)

\bibitem{Strunz03a} W. T. Strunz, F. Haake and D. Braun, Phys. Rev. A 
\textbf{67}, 022101 (2003)

\bibitem{Strunz03b} W. T. Strunz and F. Haake, Phys. Rev. A \textbf{67},
022102 (2003)

\bibitem{Lambropoulos00a} P. Lambropoulos, G. M. Nikolopoulos, T. R. Nielsen
and S. Bay, Rep.\ Prog.\ Phys.\ \textbf{63}, 455 (2000)

\bibitem{Zwanzig64a} R. Zwanzig, Physica (Amsterdam) \textbf{30}, 1109 (1964)

\bibitem{Nakajima58a} S. Nakajima, Prog. Theor. Phys. \textbf{20}, 948 (1958)

\bibitem{Dalton82a} B. J. Dalton, J. Phys. A \textbf{15}, 2157 (1982)

\bibitem{Barnett01a} S. M. Barnett and S. Stenholm, Phys. Rev. A \textbf{64}%
, 033808 (2001)

\bibitem{Shibata77a} F. Shibata, Y. Takahashi and H. Hashitsuma, J. Stat.
Phys. \textbf{17}, 171 (1977)

\bibitem{Cresser00a} J. C. Cresser, Laser Phys. \textbf{10}, 1 (2000)

\bibitem{Breuer99a} H-P. Breuer, B. Kappler and F. Petruccione, Phys. Rev. A 
\textbf{59}, 1633 (1999)

\bibitem{Strunz99a} W. T. Strunz, L. Diosi and N. Gisin, Phys. Rev. Lett. 
\textbf{82}, 1801 (1999)

\bibitem{Walls99a} M. W. Jack, M. J. Collett and D. F. Walls, Phys. Rev. A 
\textbf{59}, 2306 (1999)

\bibitem{Molmer99a} K. M{\o}lmer and S. Bay, Phys. Rev. A \textbf{59}, 904
(1999)

\bibitem{John97a} T. Quang and S. John, Phys. Rev. A \textbf{56}, 4273 (1997)

\bibitem{Imamoglu96a} P. Stenius and A. Imamo\={g}lu, Quantum Semiclass. Opt. 
\textbf{8}, 283 (1996)

\bibitem{Breuer04a} H-P. Breuer, Phys. Rev. A \textbf{70}, 01206 (2004)

\bibitem{Knight99a} E. Paspalakis, D. Angelakis and P. L. Knight, Opt. Comm. 
\textbf{172}, 229 (1999)

\bibitem{Lambropoulos97a} S. Bay, P. Lambropoulos and K. M{\o}lmer, Phys. Rev.
Lett. \textbf{79}, 2654 (1997)

\bibitem{Law00a} C. K. Law, T. W. Chen and P. T. Leung, Phys.\ Rev.\ A 
\textbf{61}, 023808 (2000)

\bibitem{Garraway97a} B. M. Garraway, Phys. Rev. A \textbf{55}, 2290 (1997)

\bibitem{Garraway97b} B. M. Garraway, Phys. Rev. A \textbf{55}, 4636 (1997)

\bibitem{Fano61a} U. Fano, Phys. Rev. \textbf{124}, 1866 (1961)

\bibitem{Barnett00a} J. Jeffers, P. Horak, S. M. Barnett and P. M. Radmore,
Phys. Rev. A \textbf{62}, 043602 (2000)

\bibitem{Privman02a} V. Privman, Mod. Phys. Lett. B \textbf{16}, 459 (2002)

\bibitem{Privman03a} V. Privman, J. Stat. Phys. \textbf{110}, 957 (2003)

\bibitem{Privman04b} D. Tolkunov and V. Privman, Phys. Rev. A \textbf{69},
062309 (2004)

\bibitem{Duan97a} L-M. Duan and G-C. Guo, Phys. Rev. A \textbf{56}, 4466
(1997)

\bibitem{Zagury01a} J. C. Retamel and N. Zagury, Phys. Rev. A \textbf{63},
032106 (2001)

\bibitem{Dalton01a} B. J. Dalton, S. M. Barnett and B. M. Garraway,\
Fortschr. Phys. \textbf{49}, 927 (2001)

\bibitem{Dalton01b} B. J. Dalton, S. M. Barnett and B. M. Garraway, Phys.\
Rev.\ A \textbf{64}, 053813 (2001)

\bibitem{Dalton02a} B. J. Dalton and B. M. Garraway, J.\ Mod.\ Opt.\ \textbf{%
49}, 947 (2002)

\bibitem{Dalton03a} B. J. Dalton and B M. Garraway, Phys.\ Rev.\ A \textbf{68%
}, 033809 (2003)

\bibitem{Nikolopoulos00a} G. M. Nikolopoulos and P. Lambropoulos, Phys.\
Rev.\ A \textbf{61}, 053812 (2000)

\bibitem{Bay98a} S. Bay and P. Lambropoulos, Opt. Comm. \textbf{146}, 130
(1998)
\end{thebibliography}
\end{document}